\documentclass[twocolumn,preprintnumbers,amsmath,amssymb,prl,superscriptaddress]{revtex4}

\usepackage{graphicx} 
\usepackage{dcolumn}  
\usepackage{colordvi}
\usepackage{color}
\graphicspath{{ps}}

\newcommand{\dz}{\ensuremath{D^0}}
\newcommand{\dzbar}{\ensuremath{\overline{D}{}^0}}

\begin{document}

\preprint{\vbox{ \hbox{   }
    \hbox{}
    \hbox{}
    \hbox{} 
}}

\title { \quad\\[0.5cm]  Recent results on $D^0$ mixing from Belle}

\author{B.~Golob}\affiliation{University of Ljubljana,
  Ljubljana, Slovenia}\affiliation{J. Stefan Institute, Ljubljana, Slovenia} 

\collaboration{The Belle Collaboration}

\date{\today}

\begin{abstract}
  We report on recent measurements of the $D^0-\overline{D}^0$ mixing
  and $CP$ violation parameters performed by the Belle
  experiment. The
  evidence for the mixing phenomena in the system of neutral $D$
  mesons, arising in the study of $D^0\to K^+K^-, \pi^+\pi^-$
  decays is presented first. Using a time dependent Dalitz analysis of $D^0\to
  K_S\pi^+\pi^-$ decays we also obtained the most precise up-to-date determination of the mass
  difference of the two $D$ meson mass eigenstates. 
The presented results are based on 540~fb$^{-1}$ of data recorded
  by the Belle detector at the KEKB $e^+e^-$ collider. We conclude
  with short prospects for the future measurements.  
\end{abstract}
\maketitle

\section{Introduction}

Measurements in the field of charmed hadrons experience a revival
in the recent years. The reason for an increased interest is twofold:
the B-factories provide for an abundant source of charmed hadrons. The
integrated luminosity ${\cal{L}}\approx 700$ fb$^{-1}$ of the KEKB
  collider \cite{KEKB} corresponds to a production of around
  $900\times 10^6$ charmed hadron pairs in a clean environment of
  $e^+e^-$ collisions. Secondly, a dual role of charm physics is
  exploited: as an experimental test ground for different theoretical
  predictions, most notably the lattice QCD, enabling in turn a more precise
  determinations of the Cabibbo-Kobayashi-Maskawa (CKM) matrix elements; and
  as a standalone field of various Standard model (SM) tests and
  searches of new physics (NP) phenomena. 

Search for the $D^0-\overline{D}^0$ mixing (a quest started soon after the
discovery of \dz mesons in 1976 \cite{goldhaber}) belongs to the latter category. It
is governed by the lifetime of $D$ mesons, $\tau=1/\Gamma$, and by
the mixing parameters $x=(m_1-m_2)/\Gamma$ and
$y=(\Gamma_1-\Gamma_2)/2\Gamma$. $m_{1,2}$ and $\Gamma_{1,2}$ denote
the masses and widths of the mass eigenstates $D_1$ and $D_2$,
respectively, 
\begin{equation}
|D_{1,2}\rangle=p|D^0\rangle\pm q |\overline{D}^0\rangle~~.
\label{eq0}
\end{equation}
$\Gamma=(\Gamma_1+\Gamma_2)/2$ is the average decay width. The mixing rate is
severely suppressed due to the small SU(3) flavor symmetry breaking
($m_s^2\approx m_{u,d}^2$) and smallness of the $|V_{ub}|$ CKM matrix
element \cite{shipsey}. Calculations based
on the effective $\Delta C=2$ Hamiltonian (i.e. contribution
of the box diagram, providing a satisfactory description of mixing in
the systems of $K^0, B_d^0$ and $B_s^0$ mesons) yield a negligible
mixing parameter magnitude $|x|\sim{\cal{O}}(10^{-5})$. Long distance
contributions to the \dz - \dzbar ~ transitions are difficult to
calculate. Approaches based on the operator product expansion
\cite{bigi1} or summation over exclusive intermediate states,
accessible to both $D^0$ and \dzbar ~\cite{falk}, result in $|x|, |y|\le 10^{-3}$ and 
$|x|, |y|\le 10^{-2}$, respectively. 

\begin{figure}[h]
  \includegraphics[width=6cm]{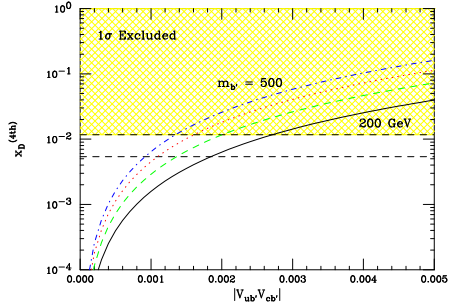}
  \caption{Relation between the mixing parameter $x$ and product of
    CKM matrix elements in model including the fourth generation down-like
    quark $b^\prime$ \cite{golowich}.}
  \label{fig0}
\end{figure}

Regardless of large uncertainties of the predictions, the SM 
probability of $D^0$ meson to oscillate into
its antiparticle before decaying, $R_M\approx (x^2+y^2)/2$, is small, at most 
of the order ${\cal{O}}(10^{-4})$. New as-yet-unobserved particles
could contribute to the
loop diagrams and thus significantly increase the value of $|x|$. 
A limit of $|x|\lesssim {\cal{O}}(10^{-2})$ would 
put significant constraints on the parameter space of
large number of NP models \cite{golowich}. As an example a possibility
of the fourth family of fermions, including a down-like $b'$
quark of mass $m_{b'}$, may be considered (Fig. \ref{fig0}). An experimental value of
$|x|\lesssim 10^{-2}$ would constrain the CKM matrix elements product
$|V_{ub'}V_{cb'}|\lesssim 3\times 10^{-3}$ for $m_{b'}\gtrsim 200$~GeV/c$^2$, a
limit much more stringent than the one following from the current CKM
matrix unitarity requirement. Similar parameter constraints can be put to 
17 out of 21 NP models considered in \cite{golowich}. The width
difference ($y$), on the other hand, is governed by $D$ decays into
physical states, where no significant deviations from the SM have
been observed up to date. However, $y$ is known to vanish in the limit
of exact SU(3)
flavor symmetry. Hence NP contributions not vanishing in this limit 
could affect the value of $y$ regardless
of their small contribution to the decay amplitudes
\cite{golowich2}. 

Beside the possible effects on the mixing parameters, NP could produce
a sizable violation of the $CP$ symmetry ($CPV$) in $D$ meson decays \cite{nir}. Within
the SM the $CPV$ is expected to be small. Of the three types of the
$CP$ violation, $CPV$ in decays, $CPV$ in mixing and in the interference
between mixing and decays, the first one is expected to be present
only in the singly Cabibbo suppressed decays. To these, beside the tree
amplitude also penguin diagrams ($c\to u\bar{q} q$) can contribute, and the existence
of at least two amplitudes of different strong and weak phase is a necessary
condition for this type of violation to occur \cite{pdg_cpv}. The weak phase difference
between the two amplitudes is $\sim \Im(V_{cd}V_{ud}^\ast
  V_{cs}V_{us}^\ast)\le 10^{-3}$, which represents a rough estimate of
  the expected $CPV$ effect \cite{shipsey}. The asymmetries due to $CPV$ in
  mixing and interference can be expressed as $A_{CP}\propto
  y\cos{\phi}$ and $\propto x
  \sin{\phi}$, respectively, where $\phi\sim 10^{-3}$ is the weak phase between
  the mixing and decay amplitudes. Hence the magnitude of these types 
  of $CPV$ is even smaller. Observation of the $CP$ violation
  an order of magnitude larger than these expectation would clearly point to
  the intervention of NP.  

\begin{figure}[h]
  \includegraphics[width=3.25cm]{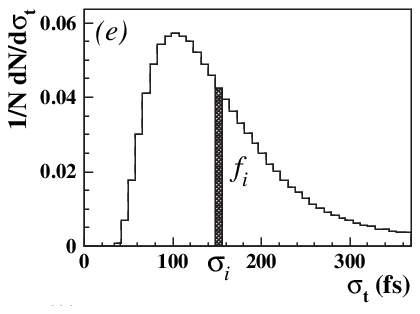}
\includegraphics[width=3.25cm]{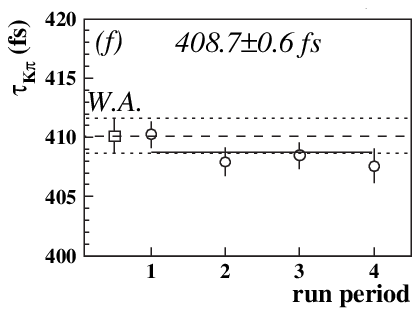}
  \caption{Left: Normalized distribution of estimated errors on the
    decay time $t$ in 
    $D^0\to K^-\pi^+$ decays. Fraction $f_i$ of events has an
    uncertainty of $\sigma_i$. Right: Measured $D^0$ lifetime from
    $D^0\to K^-\pi^+$ in different running periods. The value shown on the
    top is the average. The left-most 
    point shows the current world average of lifetime \cite{PDG}.}
  \label{fig1}
\end{figure}

The presented studies were carried out by the Belle detector, 
a general purpose full solid-angle spectrometer
\cite{belle} operating at the asymmetric $e^+e^-$ KEKB collider \cite{KEKB}. The
center-of-mass (CMS) energy of the collisions corresponds to the mass of
the $\Upsilon (4S)$, decaying to a pair of $B$ mesons. In addition to
the $B\overline{B}$ production,  
the cross section for the continuum
production of $u, d, s$ and $c$ quark pairs through a virtual photon
exchange at this energy is several times larger. In the presented
measurements the $D^0$ mesons produced 
in $e^+e^-\to c\overline{c}$ are reconstructed \footnote{Unless
  explicitly noted, mentioned processes and particles imply also the charge conjugated ones.}. Two main
features of the detector are exploited for this purpose. The 
identification of detected charged tracks is performed using a
combined information from several detector sub-modules
\cite{nakano}. An illustration of the performance can be given by the
efficiency for the charged kaon identification, $\sim 90\%$, with the
$\pi^\pm$ misidentification rate $\le 10\%$ for tracks with momenta
between 1~GeV/$c$ and 3.5~GeV/$c$. A silicon vertex detector enables
a precise determination of the decay time $t$ of short-lived particles. For
$D^0\to K^-\pi^+$ decays,   
the distribution of estimated uncertainties
on $t$ peaks at around $\tau(D^0)/3$ and has an average of
$\tau(D^0)/2$ (see Fig. \ref{fig1}, left \cite{staric}); $\tau(D^0)$ is
the 
world average value of the $D^0$ lifetime \cite{PDG}. 

\section{Measurements} 

There is a long list of measurements devoted to the \dz -
\dzbar ~ mixing from the Belle collaboration. Both, semileptonic
\cite{urban} and hadronic decays \cite{liming1} have been exploited in
the past. The most sensitive recent measurements, using decays to a
$CP$ eigenstate ($D^0\to f_{CP}$, with
$f_{CP}=K^+K^-,\pi^+\pi^-$) \cite{staric} and to a self-conjugate
final state ($D^0\to K_S\pi^+\pi^-$) \cite{liming2}, are presented in this
paper. 

There are several methods and selection criteria in common to the
presented measurements. In order to search for events where a $D^0$ 
undergoes a transition to $\overline{D}^0$ the flavor of the initially produced
neutral meson must be tagged. This is achieved by reconstruction of
decays $D^{\ast +}\to D^0\pi_s^+$ or $D^{\ast -}\to
\overline{D}^0\pi_s^-$. The charge of the characteristic low momentum
pion $\pi_s$ tags the flavor of the initially produced $D$ meson. The
energy released in the $D^\ast$ decay, 
\begin{equation}
q=M(D^\ast)-M(D^0)-m_\pi~~,
\label{eq1}
\end{equation}
has a narrow peak for the signal events and thus
helps in rejecting the combinatorial background. 
Here, $M(X)$ is used to denote the invariant mass of the $X$
decay products, and $m_X$ for the nominal mass of $X$. $D^0$ 
mesons produced in $B$ decays have different decay time distribution
and kinematic properties than the mesons 
produced in continuum. In order to obtain a sample of neutral
mesons with uniform properties we require the momentum of the reconstructed $D^\ast$ mesons
in the CMS to be larger than 2.5~GeV/c$^2$. Since the
momentum of $D^\ast$ in $B\to D^\ast X$ is kinematically constrained, this 
requirement completely rejects the $D^0$'s from the latter
source. Last but not least, the selection criteria are optimized using
the MC simulation, in order not to bias the results of the measurements. 

\subsection{Evidence for charm mixing in $D^0\to K^+K^-/\pi^+\pi^-$}

In the limit
of no $CPV$ the mass eigenstates of neutral charmed mesons, with
distinct values of lifetime $1/\Gamma_{1,2}$, are also
$CP$ eigenstates. Hence only the mass eigenstate component of $D^0$ 
with the $CP$ eigenvalue equal to the one of $f_{CP}$ contributes to 
$D^0\to f_{CP}$ decays. By measuring the lifetime of $D^0$ in decays
to $f_{CP}$ one determines the corresponding $1/\Gamma_1$ or
$1/\Gamma_2$. On the other
hand, both $CP$ states contribute in decays to non-$CP$ final states,
like $K^-\pi^+$. The measured value of the effective lifetime in the
latter process corresponds to a mixture of $1/\Gamma_1$ and 
$1/\Gamma_2$. By explicit writing of decay time rates in the presence
  of oscillations, and taking into account $|y|<<1$, one derives a
  relation between lifetimes as measured in $D^0\to f_{CP}$ and in
  decays to a mixed $CP$ final state to be \cite{bergman} 
\begin{equation}
\tau(f_{CP})=\frac{\tau(D^0)}{1+\eta_f y_{CP}}~~, 
\label{eq2}
\end{equation}
with $\eta_f=\pm 1$ denoting the $CP$ eigenvalue of $f_{CP}$. $\tau(D^0)$
represents the effective $D^0$ lifetime as measured in decays to
non-$CP$ eigenstates ($1/\Gamma$) and the relative difference of
the lifetimes is described by the parameter $y_{CP}$. 

The final states $f_{CP}=K^+K^-, \pi^+\pi^-$ are $CP$ eigenstates with
$\eta_{K^+K^-,\pi^+\pi^-}=+1$. The ratio of lifetimes measured in
these decays and in the $D^0\to K^-\pi^+$ yields the value of $y_{CP}$: 
\begin{equation}
y_{CP}=\frac{\tau(K^-\pi^+)}{\tau(f_{CP})}-1~~. 
\label{eq3}
\end{equation}
Expressed in terms of the mixing parameters, $y_{CP}$ reads
\cite{bergman} 
\begin{equation}
y_{CP}=y\cos{\phi}-\frac{1}{2}A_M\sin{\phi}~~, 
\label{eq4}
\end{equation}
with $A_M$ and $\phi$ describing the $CPV$ in mixing and interference
between mixing and decays, respectively. If for the moment 
the possibility of $CPV$ is neglected ($A_M, \phi =0$; search for the $CPV$ is described
separately in a later section), one notes that $y_{CP}=y$. The described
method of $y_{CP}$ determination has been exploited in \cite{staric}. 

\begin{figure}[h]
  \includegraphics[width=7cm]{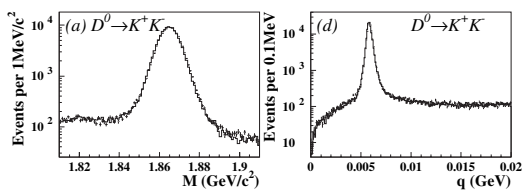}
  \caption{Left: $M(D^0)$ distribution for $D^0\to K^+K^-$ with
    $|\Delta q| < 0.80$~MeV. Right: $\Delta q$ distribution for decays
    with $M(D^0)/\sigma_M<2.3$. Full histograms represent the results of the
    fits. }
  \label{fig2}
\end{figure}

Selection of the $D^0$ candidate decays is based on $M(D^0)/\sigma_M$
(where $\sigma_M$ is the decay channel dependent resolution), $\Delta
q=q-(m_{D^\ast}-m_D-m_\pi)$ and
$\sigma_t$. Distributions of $M(D^0)$ and $q$ for $D^0\to K^+K^-$ are
shown in Fig. \ref{fig2}. 
The signal yields and purities of selected $f_{CP}$ and
$K^-\pi^+$ samples, following from the fit to the tuned simulated
samples, are given in Table \ref{tab1}. The amount of background in
all three reconstructed channels is
low.  
\begin{table}
  \caption{\label{tab1} Signal yields and purities of selected
    samples. }
  \begin{ruledtabular}
    \begin{tabular}{c c c}
Final state & Signal yield & Purity \\ \hline\hline
$K^+K^-$ & 111$\times 10^3$ & 98\% \\ \hline
$K^-\pi^+$ & 1220$\times 10^3$ & 99\% \\ \hline
$\pi^+\pi^-$ & 49$\times 10^3$ & 92\% 
    \end{tabular}
  \end{ruledtabular}
\end{table}

Final state tracks are refitted to a common $D^0$ decay vertex. The
production point is found by constraining the $D^0$ momentum vector
and $\pi_s$ from the $D^\ast$ decay to originate from the $e^+e^-$
interaction region. The proper decay time $t$ is calculated as a
projection of the vector $\vec{L}$ joining the two vertices onto the
momentum of $D^0$, $t=m_D\vec{L}\cdot\vec{p}/p^2$. 
To determine $y_{CP}$ we perform a simultaneous
binned likelihood fit to the decay time distributions in the three
decay modes, with lifetimes related by a free parameter $y_{CP}$.  

The $t$ distributions are described as a sum of the signal and
background contribution $B(t)$. The signal contribution is a convolution of
an exponential and a detector resolution function $R(t)$: 
\begin{equation}
  dN/dt = \frac{N_{\rm{sig}}}{\tau}\int e^{-t'/\tau} \cdot R(t-t')\, \mathrm{d}t' + B(t).  
\label{eq5}
\end{equation}

The composition of the resolution function is illustrated in
Fig. \ref{fig1}, left. Normalized distributions of estimated $\sigma_t$,
based on the uncertainties of the decay length determination, are plotted
for individual decay channels. In an ideal case each $\sigma_i$ value
represents a Gaussian resolution term with a weight $f_i$. Study of
the normalized residual distributions, $(t_{\rm reconstructed}-t_{\rm
  generated})/\sigma_t$, however, reveals that they cannot be described by a
single Gaussian function. They are well described by the sum of three
Gaussians, $\sum_{k=1}^{3}w_k G(t_{\rm reconstructed}-t_{\rm
  generated};\sigma_k^{\rm pull}, t_0)$, with weights $w_k$, widths
$\sigma_k^{\rm pull}$ and a common mean $t_0$. It follows that each
$\sigma_i$ represents a resolution term composed of three
Gaussians. The final parametrization of the resolution function is
thus 
\begin{equation}
  R(t-t')=\sum_{i=1}^n{f_i}\sum_{k=1}^3{w_kG(t-t';\sigma_{ik},t_0)}~~,
  \label{eq6}
\end{equation}
with $\sigma_{ik}=s_k\sigma_k^{\rm pull}\sigma_i$. The scale factors
$s_k$ are introduced to describe small differences between the
simulated and real $\sigma_k^{\rm pull}$. 

The background distribution is represented by a sum of an
exponential and $\delta$ function, convolved with the resolution
function parametrized as above. Parameters of $B(t)$ are determined
from fits to $t$ distributions of events in the $M(D^0)$ sidebands. 

Several running periods, coinciding with changes to the detector, were
identified based on the resulting $\tau(K^-\pi^+)$. For one of the
periods the resolution function (\ref{eq6}) is modified to be slightly
asymmetric in order to yield a consistent lifetime. This is achieved
by introducing a decay mode dependent difference of $t_0$'s of the
first two Gaussian terms in $R(t)$. This behaviour has been reproduced
by generating a special MC sample which includes a small additional misalignment
between the vertex detector and central drift chamber of the Belle
detector. The lifetime measured in $D^0\to K^-\pi^+$ decays shows a
good consistency among the running periods as well as with the world
average value \cite{PDG} (Fig. \ref{fig1}, right). 

\begin{figure}[h]
  \includegraphics[width=7cm]{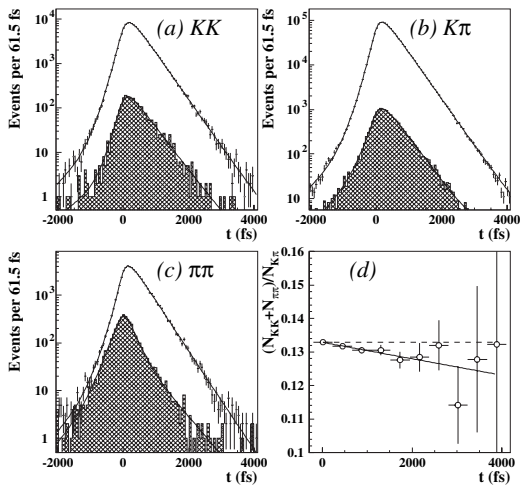}
  \caption{(a)-(c): Result of the simultaneous fit to decay time distributions in
    $D^0$ decays to individual final states. The hatched areas
    represent the contribution of backgrounds. (d): Ratio of $D^0\to
    f_{CP}$ and $D^0\to K^-\pi^+$ decay time distributions. The slope
    visualize the difference of effective lifetimes.}
  \label{fig3}
\end{figure}

Simultaneous fits to decay time distributions of $K^+K^-, K^-\pi^+$
and $\pi^+\pi^-$ were performed for individual running periods and the
resulting $y_{CP}$ values averaged to obtain the final result. Fits
are presented in Fig. \ref{fig3}(a)-(c) by summing the data points and the
fit function values. The agreement of the fit function with the data
is excellent, $\chi^2/n.d.f=312/289$. The same is true for all
individual fits as well. The final value obtained is 
\begin{equation}
y_{CP}=(1.31\pm 0.32(\rm stat.)\pm 0.25(\rm syst.))\%~~.
\label{eq7}
\end{equation}

The largest systematic uncertainties follow from the assumption of
equal $t_0$ for different decay channels (estimated by relaxing this
constraint), possible deviations of acceptance dependence on decay
time from a constant (estimated by a fit to the generated $t$
distribution of reconstructed MC events) and variation of
selection criteria (effect estimated using high statistics MC
samples). 

The resulting $y_{CP}$ deviates from the null value by more than 3
standard deviations (more than 4 standard deviations considering the 
stat. error only) and represents a clear evidence of
$D^0-\overline{D}^0$ mixing, regardless of possible $CPV$. The
difference of lifetimes is made
visually observable by ploting the ratio of decay time distributions for decays to
$f_{CP}$ and $K^-\pi^+$ in Fig. \ref{fig3}(d).

\subsection{Measurement of charm mixing parameters in $D^0\to
  K_S\pi^+\pi^-$}

\begin{figure}[h]
  \includegraphics[width=5cm]{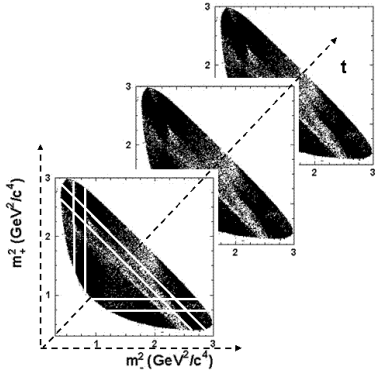}
  \caption{Illustration of the decay time dependent Dalitz
  analysis. Contributions of different intermediate states (sketched
  by white lines) can be
  identified by analysis of the Dalitz distribution. Different types
  of decays exhibit a different decay time propagation. By studying
  the time evolution of the Dalitz distribution one effectively
  determines the $t$ distribution of different types of decays and by this
  the mixing parameters $x$ and $y$.}
  \label{fig4}
\end{figure}

To a hadronic multi-body final state several intermediate resonances
can contribute. In a specific example of the self-conjugated mode $D^0\to K_S\pi^+\pi^-$
contributions from Cabibbo favored decays (e.g. $D^0\to K^{\ast
  -}\pi^+$), doubly Cabibbo suppressed decays (e.g. $D^0\to K^{\ast
  +}\pi^-$) and decays to $CP$ eigenstates (e.g. $D^0\to \rho^0 K_S$)
are present. Individual contributions can be identified by analyzing
the Dalitz distribution of the decay (see Fig. \ref{fig4}). 

Different types of intermediate states exhibit also a specific time
evolution (due to their specific superposition from the $D$ mass
eigenstates). While the decay time distribution of $CP$ eigenstates
depends on the parameter $y$, the $t$ evolution of doubly Cabibbo
suppressed decays depends on $x^\prime = x\cos{\delta} + y\sin{\delta}$
and $y^\prime = y\cos{\delta} - x\sin{\delta}$, where $\delta$ is a
strong phase difference between these and the corresponding Cabibbo
favored decays. Since in decays to $K_S\pi^+\pi^-$ both types
interfere  
it is possible to disentangle the relative
phase by performing a fit to the Dalitz distribution. This in turn
enables a direct determination of the mixing parameters $x$ and $y$
instead of their rotated versions $x^\prime$ and $y^\prime$. The
method was successfully exploited in \cite{liming2}. 

The two Dalitz variables are defined as $m_-^2=M^2(K_S\pi^-)$ and
$m_+^2=M^2(K_S\pi^+)$. Decay time dependent matrix element for $D^0\to
  K_S\pi^+\pi^-$ decay, where the initially produced $D$ meson is a
  $D^0$, is written as 
\begin{align}
\nonumber
&{\cal{M}}(m_-^2,m_+^2,t)=\langle K_S\pi^+\pi^-|D^0(t)\rangle=&\\
\nonumber
&=\frac{1}{2}{\cal{A}}(m_-^2,m_+^2)\bigl[e^{-i\lambda_1
      t}+e^{-i\lambda_2 t}\bigr]+&\\
&+\frac{1}{2}\overline{\cal{A}}(m_-^2,m_+^2)\bigl[e^{-i\lambda_1 t}-
  e^{-i\lambda_2 t}\bigr]&~~.
\label{eq8}
\end{align}
In the above expression ${\cal{A}}(m_-^2,m_+^2)$ and $\overline{\cal{A}}(m_-^2,m_+^2)$ are the
instantaneous amplitudes for $D^0$ and $\overline{D}^0$
decays. The dependence on the mixing parameters arises upon squaring
the matrix element in which
$\lambda_{1,2}=m_{1,2}-i\Gamma_{1,2}/2$. $\overline{\cal{M}}(m_-^2,m_+^2,t)$
describing the decay of an initially produced $\overline{D}^0$ is
written in an analogous form. Neglecting $CPV$ one finds
${\cal{M}}(m_-^2,m_+^2,t)=\overline{\cal{M}}(m_+^2,m_-^2,t)$. 

Amplitudes for $D$ decays are parametrized in the isobar model as a
sum of Breit-Wigner resonances and a constant non-resonant term: 
\begin{equation}
{\cal{A}}(m_-^2,m_+^2)=\sum_r a_r e^{i\phi_r}B_r(m_-^2,m_+^2)+a_{\rm
  NR}e^{i\phi_{\rm NR}}~~.
\label{eq9}
\end{equation}
Functions $B_r$ are products of Blatt-Weisskopf form factors and
relativistic Breit-Wigners \cite{cleo_Dalitz}. The described signal
distribution is convolved by the detector mass resolution function (for
$\pi^+\pi^-$ invariant mass only) and multiplied by $(m_-^2,m_+^2)$
dependent efficiency. The expected decay time distribution is
convolved with a resolution function described by a sum of three
Gaussians with a common mean. The mean and scale factors for the widths of the
resolution function are free parameters of the fit.

\begin{figure}[h]
  \includegraphics[width=3.5cm]{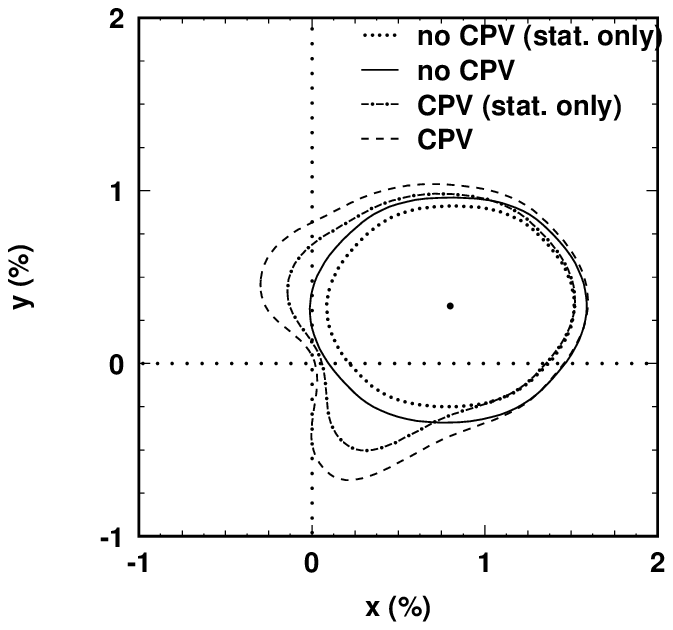}\includegraphics[width=3.5cm]{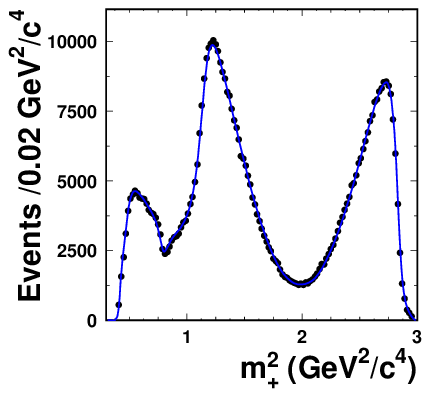}
  \includegraphics[width=7cm]{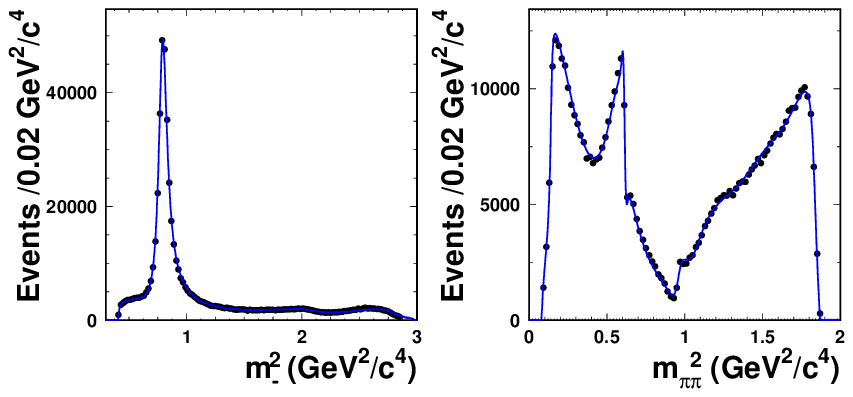}
  \caption{Top left: 95\% C.L. region for parameters $x$ and $y$ as obtained in 
$D^0\to K_S\pi^+\pi^-$ decays. Top right and bottom: Projections of
    the Dalitz distribution to squares
  of two-particle invariant masses. Full line is the result of the
  fit. Note that in the plots $m_\pm^2$ corresponds to
  $M^2(K_S\pi^\pm)$ for $D^0$ decays and to $M^2(K_S\pi^\mp)$ for
  $\overline{D}^0$ decays.}
  \label{fig5}
\end{figure}

The isolation of signal is based on $M(D^0)$ and $q$ variables,
described in the previous section. Selected sample of decays used for
the measurement consists of $534\times 10^3$ signal decays with a
purity of 95\%. Fractions of individual backgrounds are
obtained from the two-dimensional fit of $M(D^0)$ and $q$
distributions. 

The Dalitz distribution of combinatorial background
(4\%) is
obtained from events in the $M(D^0)$ sidebands. This probability
density function (p.d.f.) is multiplied by a sum of an exponential and $\delta$
functions to represent the $t$ distribution, and convolved with the
same resolution function as for the signal. The background with a
true $D^0$ and a random slow pion represents 1\% of the sample. The
p.d.f. of this background is the same as the one of the signal. 

An unbinned likelihood fit is performed to distribution of events in
the signal region. Results of the fit in which we neglect possible
$CPV$ are projected to the Dalitz variables in Fig. \ref{fig5}. 
The Dalitz model which includes 18
intermediate states represents a good description of the data. 

\begin{figure}[h]
  \includegraphics[width=5cm]{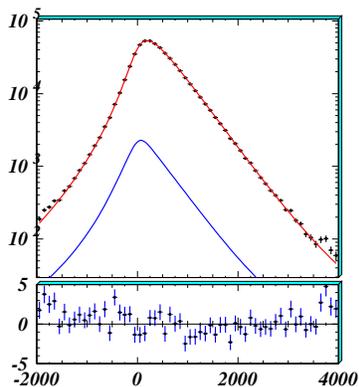}
  \caption{Top: Decay time distribution of selected $D^0\to
   K_S\pi^+\pi^-$. 
   Full line is the result of the fit. Contribution of background is
   presented by the lower line. Time scale is given in fs. Bottom:
   Residuals between the data and
   the fitting function. }
  \label{fig6}
\end{figure}

The effective value of the $D^0$ lifetime following from the fit,
$\tau(D^0)=(409.9\pm 0.9)$~fs, is in good agreement with the world
average value \cite{PDG}. The decay time projection of the fit is
presented in Fig. \ref{fig6}. 

Finally, the result for mixing parameters is 
\begin{align}
&x=(0.80\pm 0.29({\rm stat.}){+0.13\atop -0.16}({\rm syst.}))\%& \\
&y=(0.33\pm 0.24({\rm stat.}){+0.10\atop -0.14}({\rm syst.}))\%
\label{eq10}
\end{align}

The systematic uncertainties are divided into two categories:
uncertainties related to the Dalitz model, and others. The former are 
estimated by repeating the fit with the K-matrix parametrization of
the scalar resonances, and by estimating possible biases in the ratios
of doubly Cabibbo suppressed and Cabibbo favored decays using the
simulation. The largest uncertainty in the latter category arises from
the variation of the selection on the $D^\ast$ CMS momentum and from
the assumption of factorization of the Dalitz and decay time distributions
of combinatorial background (estimated by using Dalitz distribution of
combinatorial background from different $t$ intervals). 

The resulting value of the mass difference $x$ is the most precise
measurement of this parameter, improving the precision of the
previous measurement \cite{cleo_x} by an order of magnitude. 

\subsection{Search for $CPV$}

The value of $y_{CP}$ measured in $D^0\to f_{CP}$ depends on the parameters
describing $CP$ violation (eq. (\ref{eq4})). The definition of the
parameters is \cite{bergman}  
\begin{align}
\nonumber
&\frac{|q|^2}{|p|^2}\equiv 1+A_M& \\
&\frac{q}{p}\frac{{\cal{A}}(\overline{D}^0\to
    K^+K^-)}{{\cal{A}}(D^0\to K^+K^-)}\equiv
-\frac{|q|}{|p|}e^{i\phi}~~, 
\end{align}
where $A_M\ne 0$ is a sign of the $CPV$ in mixing and $\phi\ne 0$ of the
$CPV$ in the interference between mixing and decay. 
In $D$ decays to a $CP$ eigenstate 
one can define a $CP$ asymmetry: 
\begin{equation}
A_\Gamma=\frac{\tau(\overline{D}^0\to f_{CP})-\tau(D^0\to f_{CP})}
{\tau(\overline{D}^0\to f_{CP})+\tau(D^0\to f_{CP})}~~, 
\label{eq11}
\end{equation}
which can be expressed in terms of fundamental parameters as  
\begin{equation}
A_\Gamma=\frac{1}{2}A_M y\cos{\phi} - x\sin{\phi}~~.
\label{eq12}
\end{equation}
Hence by separately measuring the lifetime of $D^0$ and
$\overline{D}^0$ tagged decays, we measure \cite{staric}
\begin{equation}
A_\Gamma=(0.01\pm 0.30({\rm stat.})\pm 0.15({\rm syst.}))\%~~. 
\label{eq13}
\end{equation}
Systematic uncertainty receives similar contributions as in the case
of $y_{CP}$ measurement. 
The $CPV$ in mixing and interference is thus not observed with a
sensitivity of $\sim 0.35\%$. 

To search for $CPV$ in $K_S\pi^+\pi^-$ decays a more general fit than the one described
in the previous section is performed \cite{liming2}. In addition to previous
parameters we allow for $|q/p|\ne 1$ and $\phi\ne 0$. To check for a
possibility of $CPV$ in decays the parameters $a_r$ and $\phi_r$ of
Eq. (\ref{eq9}) are allowed to be different for $D^0$ and
$\overline{D}^0$ decays. The resulting Dalitz parameters are
consistent for the two samples and no sign of $CPV$ in decays is
observed. Results of the consequent fit assuming no direct $CPV$ are 
\begin{align}
\nonumber
&\frac{|q|}{|p|}=0.86 {+0.30\atop -0.29}({\rm stat.}) {+0.10\atop
    -0.09}({\rm syst. }) &\\
&\phi=(-0.24 {+0.28\atop -0.31}({\rm stat.}) \pm 0.09({\rm
    syst.}))~~{\rm rad}~~.
\label{eq14}
\end{align}
Also this measurement shows no evidence of $CPV$. 

The 95\% confidence level region in $x, y$ plane, following from the
no-$CPV$ and $CPV$ allowed fits to $D^0\to K_S\pi^+\pi^-$ decays, is
shown in Fig. \ref{fig5}, top left. It should be noted that if the $CPV$
parameters are left free in the fit, the solution $(-x,-y,arg(q/p)+\pi)$
is an equally probable solution as the $(x,y,arg(q/p))$. In terms of
Fig. \ref{fig5} this means that contours reflected over the $(0,0)$ point
also represent an allowed region of parameters space. A peculiar shape
of the $CPV$ allowed contour in the vicinity of the $(0,0)$ point is a
consequence of no possible $CPV$ in mixing or in interference between
mixing and decays when $x, y=0$ (in rest of the region the sensitivity of
measurement is spread among two mixing and two $CPV$ parameters; close
to $x, y=0$ the sensitivity of the measurement is mainly to $x$ and
$y$ and thus the likelihood function becomes steeper). 

\begin{figure}[h]
  \includegraphics[width=7cm]{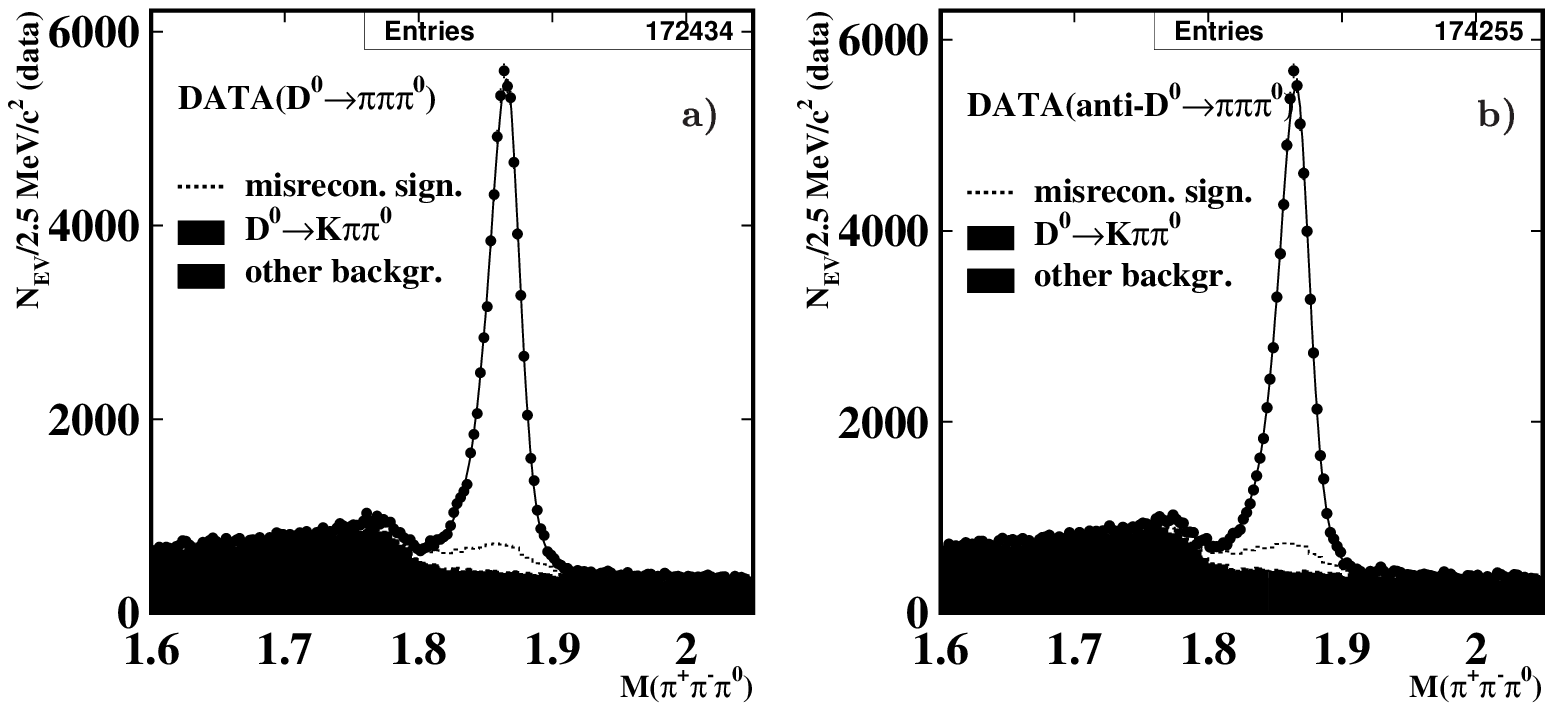}
  \caption{Fits to $M(D^0)(M(\overline{D}^0))$ distributions of
    $D^0(\overline{D}^0)\to \pi^+\pi^-\pi^0$ decays.} 
  \label{fig8}
\end{figure}

Belle also obtained a preliminary result of $t$- and Dalitz
plane-integrated 
$CPV$ search in $D$ meson decays to $\pi^+\pi^-\pi^0$. The
amount of signal and backgrounds in the selected sample is determined by a
fit to $M(D^0)$ and $M(\overline{D}^0)$ distributions, shown in
Fig. \ref{fig8}. 

The number of $D$ decays is calculated from the Dalitz distribution of
events by subtraction of the background. Background distribution is
determined using MC simulation and the uncertainty due to the
modelling is included in the systematic error. The background
subtracted yield is corrected for the
efficiency in bins of Dalitz plane. By performing the described calculation
separately for $D^0$ and $\overline{D}^0$ tagged decays, we obtain 
\begin{align}
\nonumber
&A_{CP}=\frac{\Gamma(D^0\to \pi^+\pi^-\pi^0) -
  \Gamma(\overline{D}^0\to \pi^+\pi^-\pi^0)}{\Gamma(D^0\to
  \pi^+\pi^-\pi^0) + \Gamma(\overline{D}^0\to \pi^+\pi^-\pi^0)} =&\\
 &=(0.43\pm 0.41({\rm stat.}) \pm 1.31({\rm syst.}))\%&~~.
\label{eq15}
\end{align}

\section{Outlook and summary}

\begin{figure}[h]
  \includegraphics[width=6cm]{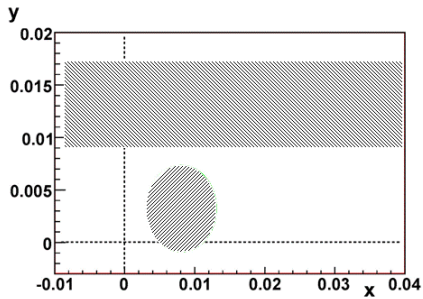}
  \caption{68\% confidence level regions of $x$ and $y$ arising from
    the lifetime measurements in $D^0\to K^+K^-,~\pi^+\pi^-$ decays and time
    dependent Dalitz analysis of $D^0\to K_S\pi^+\pi^-$.} 
  \label{fig9}
\end{figure}

Belle has recently presented a first evidence of $D^0$ mixing in
decays of charmed mesons to $CP$ eigenstates \cite{staric} and the most precise
measurement of the mass difference in the neutral charmed meson
system \cite{liming2}. The 68\% confidence regions of mixing
parameters $x$ and $y$ arising from the two measurements are presented
in Fig. \ref{fig9}. 

\begin{figure}[h]
  \includegraphics[width=6cm]{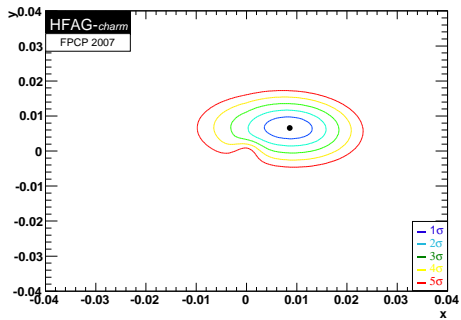}
  \caption{Contours of $n~\sigma$ 2-dimensional allowed regions for $x$
    and $y$ from the world average of measurements assuming no $CPV$.} 
  \label{fig10}
\end{figure}

The charm subgroup of the Heavy Flavor Averaging Group \cite{HFAG} has
at the time of the conference presented world averages of measurements
in the field of charm mixing and $CPV$. By summation of the likelihood
curves depending on various observables (and thus accounting for
non-Gaussian distribution of some experimental uncertainties) the $n~\sigma~(n=1-5)$
2-dimensional contours in $(x,y)$ plane are presented in
Fig. \ref{fig10}. 

The shape of the contours is a non-trivial result of summing
likelihoods of specific shapes. Especially the constraints arising from
the $D^0\to K^+\pi^-$ decays, when rotated in $(x,y)$ plane due to
the uncertainty in knowledge of the strong phase difference $\delta$,
exhibit an almost circular region of increased values around the $(0,0)$
point.

The average central values are found to be 
\begin{align}
\nonumber
&x=(0.87\pm {0.30\atop 0.34})\%&\\
&y=(0.66\pm {0.21\atop 0.20})\%&
\label{eq16}
\end{align}
The no-mixing point $(x,y)=(0,0)$ is excluded at more than $5$
standard deviations. 

Due to a larger
number of free parameters the method of averaging experimental results which allow for a
possibility of $CPV$ is a $\chi^2$ minimization. Nevertheless the results for the mixing
parameters are almost unchanged, and $CPV$ parameters are consistent
with no violation of the $CP$ symmetry:
\begin{align}
\nonumber 
&x=(0.84\pm {0.32\atop 0.34})\%&\\ 
\nonumber 
&y=(0.69\pm0.21)\%&\\ 
\nonumber 
&\frac{|q|}{|p|}=0.88\pm {0.23\atop 0.20}&\\
&\phi=(-0.09\pm {0.17\atop 0.19})~{\rm rad}&
\label{eq17}
\end{align}

The
world averages are dominated by results from the existing B-factories, 
with a significant contribution from Cleo-c and
Tevatron \cite{other_lp}. The results presented in this paper were obtained by the Belle
collaboration using around one half of the expected full data set. The
experimental errors are mainly dominated by statistical 
uncertainties and will thus improve in the next year. However, considering the
uncertainty of the SM predictions as well as the small $CPV$ expected
within, it is unlikely that the full range of high scientific interest
in the measurements of the charm mixing will be fulfilled by the end
of the data taking of Belle and BaBar. It is thus instructive to make
an attempt of predicting the accuracy that may become available
at some of the future experiments. In Table \ref{tab2} the expected
one standard deviation errors on the key parameters are given for
an average of measurements to be performed at the proposed Super-B
factory. The values are estimated by scaling the current Belle
statistical sensitivity and an educated guess on possible improvements
of systematic uncertainties. Two values of expected integrated
luminosity are considered; the lower one represents a modest data sample that
could be collected at the Super-B factory while the higher one is an
ultimate goal. The $\sigma$'s shown are of course to be taken with a grain of salt
since the systematic errors which are difficult to estimate are
important if not the dominating part of the total error. Nevertheless they 
are worth presenting, if for nothing else then for an easy-to-remember
pattern. 

\begin{table}
  \caption{\label{tab2} Expected one standard deviation errors on the
    measurements of charm mixing and $CPV$ parameters, to be performed at the Super-B factory.}
  \begin{ruledtabular}
    \begin{tabular}{c c c}
Parameter & ${\cal{L}}=5$~ab$^{-1}$ & ${\cal{L}}=50$~ab$^{-1}$  \\ \hline\hline
$\sigma(x)$ & 0.1\% & 0.07\% \\ \hline
$\sigma(y)$ & 0.1\% & 0.07\% \\ \hline
$\sigma(A_M)$ & 0.1 & 0.07 \\ \hline
$\sigma(\phi)$ & 0.1~rad & 0.07~rad \\
    \end{tabular}
  \end{ruledtabular}
\end{table}

The expected sensitivities can be compared to expectations from the 
LHCb experiment scheduled to start data taking in the coming year. With
${\cal{L}}=10$~fb$^{-1}$ (expected to be collected in around 5 years
of running at the nominal luminosity) one hopes for $\sigma(x^{\prime
  2})\approx 6\times 10^{-5}$, $\sigma(y^\prime)\approx 0.9\times
10^{-3}$ and $\sigma(y_{CP})\approx 0.05\%$ \cite{lhcb}. These
estimates can be roughly placed in between ${\cal{L}}=5$~ab$^{-1}$ and
${\cal{L}}=50$~ab$^{-1}$ expectations from the Super-B factory. 

Considering the results presented in this and other charm mixing
related papers submitted to the conference, it is fair to say that the
year 2007 (31 years after the $D^0$ meson discovery) was the
year of experimental confirmation of the $D^0$
mixing (to be compared to the time span of 6 years between similar
observations in $K^0$ system, 4 years in $B_d^0$ system and 14 years
in the case of $B_s^0$ mesons). At the moment we are facing a somewhat
rare situation of 
experimental evidence without an accurate theoretical guidance
on whether the phenomenon is entirely due
to the SM physics or not. The largest interest determining the
work ahead lies in a precise determination of $x$ and search for the
$CP$ violation in the charm sector. For it is in this field of
measurements where we can expect an answer to the above question. 
Observation of $CPV$ effects at the existing
facilities would be a sign of NP. A proposed Super-B factory would
enable searches for the $CP$ asymmetries to the $10^{-4}$ level, a
range covered by the current SM expectations and thus an
interesting area of search for NP to appear.

\end{document}